\documentclass[11pt]{article}
\usepackage{latexsym}
\begin{document}
\newcommand{\roughly}[1]%
       
\newcommand{\PSbox}[3]{\mbox{\rule{0in}{#3}
\includegraphics{#1}\hspace{#2}}}
\newcommand\lsim{\roughly{<}}
\newcommand\gsim{\roughly{>}}
\newcommand\CL{{\cal L}}
\newcommand\CO{{\cal O}}
\newcommand\half{\frac{1}{2}}
\newcommand\beq{\begin{eqnarray}}
\newcommand\eeq{\end{eqnarray}}
\newcommand\eqn[1]{\label{eq:#1}}
\newcommand\intg{\int\,\sqrt{-g}\,}
\newcommand\eq[1]{eq. (\ref{eq:#1})}
\newcommand\meN[1]{\langle N \vert #1 \vert N \rangle}
\newcommand\meNi[1]{\langle N_i \vert #1 \vert N_i \rangle}
\newcommand\mep[1]{\langle p \vert #1 \vert p \rangle}
\newcommand\men[1]{\langle n \vert #1 \vert n \rangle}
\newcommand\mea[1]{\langle A \vert #1 \vert A \rangle}
\newcommand\bi{\begin{itemize}}
\newcommand\ei{\end{itemize}}
\newcommand\be{\begin{equation}}
\newcommand\ee{\end{equation}}
\newcommand\bea{\begin{eqnarray}}
\newcommand\eea{\end{eqnarray}}

\def\Dsl{\,\raise.15ex \hbox{/}\mkern-12.8mu D}
\newcommand\Tr{{\rm Tr\,}}
\thispagestyle{empty}
\begin{titlepage}
\begin{flushright}
CALT-68-2352\\
\end{flushright}
\vspace{1.0cm}
\begin{center}
{\LARGE \bf  The Long Range Gravitational Potential Energy Between Strings}\\ 
~\\
\bigskip\bigskip
{ Margaret E. Wessling and Mark B. Wise} \\
~\\
\noindent
{\it\ignorespaces
          
\bigskip    California Institute of Technology, Pasadena CA 91125\\

}
\bigskip
\end{center}
\vspace{1cm}
\begin{abstract}
We calculate the gravitational potential energy between infinitely long parallel strings with tensions $\tau_1$ and $\tau_2$. Classically, it vanishes, but at one loop, we find that the long range gravitational potential
energy per unit length is $U/L=  24G_N^2\tau_1\tau_2/(5 \pi a^2)$ + ..., where $a$ is the separation between the strings, $G_N$ is Newton's constant, and we set $\hbar =c=1$. The ellipses represent terms suppressed by more powers of $G_N \tau_i$. Typically, massless bulk fields give rise at one loop to a long range potential between $p$-branes in space-times of dimension $p+2+1$. The contribution to this potential from bulk scalars is computed for arbitrary $p$ (strings correspond to $p=1$) and in the case of three-branes its possible relevance for cosmological quintessence is commented on.

\end{abstract}
\vfill
\end{titlepage}

In classical 2+1 dimensional general relativity, a point mass at rest does not result in a curved space-time away from the location of the particle. Instead, the space-time remains flat, but with a deficit angle cut out; the size of that angle is proportional to the mass of the particle \cite{djt}. This corresponds to a curvature singularity at the location of the particle. Hence in $2+1$ dimensional space-time there is no classical force between two point masses. Similarly, in $3+1$ dimensional general relativity, an infinitely long straight string, characterized only by its tension, leaves the exterior space-time flat, and the classical force between two
parallel infinitely long straight strings vanishes \cite{v}. The main purpose of this paper is to calculate the leading quantum mechanical long range force, or, equivalently, potential energy, between such strings. Towards the end of this paper, we will also consider contributions to the long range force that would arise if, in addition to the massless graviton, there were
a massless scalar in the bulk. We then briefly discuss the generalization of this to other co-dimension two objects ({\it i.e.} $p$-branes in $p+2+1$ dimensional space-time). In models with two large extra dimensions, this potential between three-branes may be relevant for cosmological quintessence \cite{cds}.

The action for the two string system is taken to be
\be
\label{eq1} 
S=S_b+S_1+S_2.
\ee
The bulk action, $S_b$, is the usual Einstein-Hilbert action
\be
\label{bulk}
S_b=-2M^{(n-2)}\int d^nx~\sqrt g~ R,
\ee
where $R$ is the curvature scalar, and Newton's constant $G_N$ is related to the
mass $M$ by $G_N=1/(32 \pi M^2)$. Even though the long range potential is finite, it is convenient to regulate the theory using dimensional regularization, and in equation (\ref{bulk}) $n=4-\epsilon$. For the two string actions, $S_i$, we take
\be
\label{brane}
S_i=-\tau_i \int d^nx~ \sqrt{ g^{(i)}}~ \delta^{(2)} (\vec x-\vec x_i),
\ee
where $g^{(i)}$ is the induced metric on the world-sheet of string $i$. Note that in $n$ dimensions the string world-sheets have dimension $n-2$ so they are still co-dimension two objects. We have chosen to align the strings along the $1$ axis; the separation between the two strings is $\vec a=\vec x_1-\vec x_2$. Indices that go over the 4 space-time coordinates $0,1,2,3$ ($n$ space-time coordinates in $n$ dimensions) are denoted by capital Roman letters; those that just go over the 2 space-time coordinates of the string world-sheet $0,1$ are denoted by Greek letters. Finally, indices that take on values in the two spatial directions perpendicular to the strings are denoted by lower case Roman letters, and vectors in the $2,3$ plane are denoted with arrows. We align the local space-time coordinates on the string world surfaces with those of the bulk space-time, so the components of the induced metric tensor are the same as those of the bulk metric but restricted to the $0,1$ values of the indices, {\it i.e.}  $g^{(i)}_{\alpha \beta}=g_{\alpha \beta}$.

Expanding the gravitational field as\footnote{Here $\eta=diag[-1,1,1,1]$ is the usual flat space-time metric tensor.} 
\be
g_{MN}=\eta_{MN} + h_{MN}/M^{(n /2 -1)},
\ee
we determine the leading quantum contribution to the potential between the two strings by computing one-loop Feynman diagrams with vertices that follow from the action in equation (\ref{eq1}). This is similar to the computation of the quantum correction to the Newtonian potential between point masses\footnote{ There is some ambiguity in precisely how the potential is defined. This issue is less severe for strings since the classical potential vanishes.} in four space-time dimensions \cite{dhl}. The main difference between the string
and point mass cases is that for strings the classical force vanishes; hence our computation gives the leading contribution to the force instead of a small correction.  

't Hooft and Veltman calculated the infinite part of the one-loop gravitational effective action \cite{hv}. We adopt the same background field gauge fixing, so the momentum space propagator for the canonically normalized graviton field is
\be
D_{AB,CD}(x-y)=P_{AB,CD}D(x-y), 
\ee
where $D(x-y)$ is the usual scalar propagator with Fourier transform $D(q)=-i/(q^2-i \epsilon)$ and 
\be
P_{AB,MN}={1 \over 2}[\eta_{AM} \eta_{BN} +\eta_{AN} \eta_{BM}-{2 \over n-2}\eta_{AB} \eta_{MN}].
\ee
It is convenient to use their gauge fixing convention, because then the contribution of some of the Feynman diagrams to
the quantum force between strings can be deduced from their work. Unless explicitly stated otherwise, indices on $h$ and $P$ are raised and lowered with the flat space metric tensor $\eta$. 

\begin{figure} [h] 
\PSbox{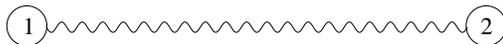 hoffset=110 voffset=10 hscale=40 vscale=40}{6.8 in}{1.0 in}
\caption{Classical contribution to the potential.  The numbers 1 and 2 represent the two string world-sheets.}
\label{fig:fig1}
\end{figure}

In this paper, we treat the tensions as small compared with $M^2$, and only keep the terms in the potential proportional to the product of the two tensions $\tau_1 \tau_2$, neglecting terms suppressed by additional powers of $G_N\tau_i$.  Using perturbation theory, it is easy to understand why the part of the classical potential proportional to $\tau_1 \tau_2$ vanishes. It comes from the Feynman diagram in Figure \ref{fig:fig1}. Using 
\be
\sqrt {g^{(i)}}=1+h_\alpha^\alpha/2M^{(n/2-1)}+ h_{\alpha}^{\alpha}~h_{\beta}^{\beta}/8M^{(n-2)}-h_{\beta}^{\alpha}~h_{\alpha}^{\beta}/4M^{(n-2)}+... , 
\ee
it follows that this diagram is proportional to $P_{\alpha}^ {\alpha},^{\mu}_{ \mu}=\eta^{\alpha \beta} \eta^{\mu \nu}P_{\alpha \beta,\mu \nu}=0$.

In background field gauge, one decomposes the graviton field into quantum and classical pieces: $h= \bar h + \tilde h$, where the bar denotes the classical part and the tilde the quantum part. The leading quantum correction occurs at one-loop. The quantum fields are contracted to make the propagators that occur in the loop; the classical fields are contracted for the other propagators.  (In the figures, classical gravitons are drawn as wavy lines, quantum gravitons as curly lines.) 

\begin{figure}
\PSbox{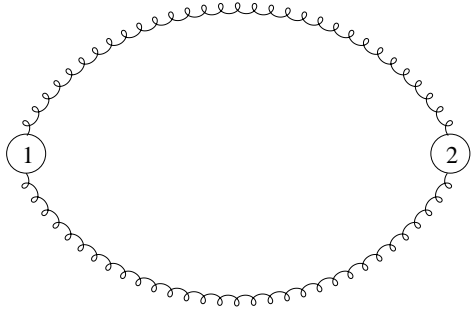 hoffset=130 voffset=10 hscale=40
vscale=40}{6.8in}{2.0in}
\caption{Feynman diagram that determines the one-loop contribution to the potential from terms localized on the string
 world-sheets that are quadratic in the graviton field.}
\label{fig:fig2}
\end{figure}

First we consider Figure \ref{fig:fig2}. The coupling of the gravitons to the string world-sheet comes from the quadratic terms in expansion of the square root of the induced metric in $h$. The contribution to the effective action that results from this Feynman diagram is

\be
\label{act1}
i\Delta S_{eff}=-{\tau_1 \tau_2 \over M^4}\left[{P^{\alpha}_{\alpha},^{\beta}_{\beta}P^{\lambda}_{\lambda},^{\delta}_{\delta} \over 32}-{P^{\alpha}_{\alpha},^{\lambda}_{\beta}P^{\delta}_{\delta},^{\beta}_{\lambda}\over 8}+{P^{\alpha}_{\beta},^{\lambda}_{\delta}P^{\beta}_{\alpha},^{\delta}_{\lambda}\over 8}\right]\int d^2x_1d^2x_2D(x_1-x_2)^2.
\ee
Equation (\ref{act1}) is evaluated using 
\be
\int d^2x_1d^2x_2D(x_1-x_2)^2=-i{LT \over a^2}\int{d^2k \over (2\pi)^2}{K_0(k)^2 \over (2\pi)^2}=-i{LT \over a^216 \pi^3},
\ee
where $K_0(k)$ is a Bessel function of imaginary argument, the integrals go over the world-sheets of the two strings, and
\be
P^{\alpha}_{\alpha},^{\beta}_{\lambda}=0, ~~~~~~~~~~~~~P^{\alpha}_{\beta},^{\lambda}_{\delta}P^{\beta}_{\alpha},^{\delta}_{\lambda}=2.
\ee
The effective action can be interpreted as minus the potential energy times the time, $\Delta S_{eff}=-\Delta U~T$.
Putting these results together, we find that the contribution to the potential energy per unit string length from this diagram is
\be
\Delta U/L={\tau_ 1\tau_2 \over 16 \pi^3 a^2M^4}\left(-{1 \over 4}\right).
\ee

\begin{figure} [h]
\PSbox{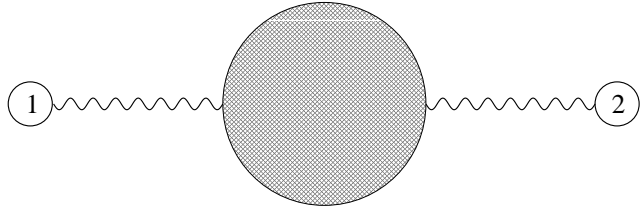 hoffset=100 voffset=10 hscale=45
vscale=45}{6.8in}{1.5in}
\caption{Feynman diagrams that give the contribution to the potential from gravitational self interactions. The shaded circle includes gravitons and ghosts in the loop.}
\label{fig:fig3}
\end{figure}

Next consider the diagrams in Figure \ref{fig:fig3}. 't Hooft and Veltman  \cite{hv} found that the divergent part of the one-loop gravitational effective action for pure Einstein gravity is
\be
\label{div}
S_{1loop}^{eff}=-{M^{n-4} \over 8 \pi^2 (n-4)}\int d^n x \sqrt {\bar g}\left({1 \over 120} \bar R^2 +{7 \over 20} \bar R_{AB} \bar R^{AB}\right).
\ee

The effective one-loop action is constructed from the classical metric $\bar g_{AB}= \eta_{AB} +\bar h_{AB}/M^{(n/2-1)}$, and indices in equation (\ref{div}) are raised and lowered with this metric.
From this effective action we can deduce the insertion appropriate for the shaded circle in Figure \ref{fig:fig3} by expanding out the curvature tensor to linear order in the gravitational field.  We get
\be
\label{r2}
\bar R^2=\left[(\partial ^2 \bar h^L_L)(\partial^2 \bar h^M_M)-2 (\partial^2 \bar h^L_L)( \partial_G \partial_E \bar h^{GE})+(\partial_K \partial_N \bar h^{KN})( \partial_G \partial_E \bar h^{GE})\right]/M^{(n-2)},
\ee
and
\bea
\label{ri2}
&& \bar R_{MK} \bar R^{MK}=\left[{1 \over 4}(\partial_K \partial_M \bar h^L_L)( \partial^K \partial ^M \bar h^N_N)-(\partial_K \partial_M \bar h^L_L)(\partial ^K \partial^N \bar h_N^M) +{1 \over 2} (\partial _K \partial_M \bar h^L_L)( \partial^2 \bar h^{KM} )\right.
\nonumber \\
&& +{1 \over 2}(\partial_K \partial_L \bar h^{LM})( \partial^K \partial^N \bar h_{MN}) +{1 \over 2} (\partial_K \partial_L \bar h^{LM})(\partial_N \partial _M \bar h^{KN})- (\partial_K \partial_L \bar h^L_M)( \partial^2 \bar h^{KM} )\nonumber \\
&&\left. +{1 \over 4 }(\partial^2 \bar h_{KM})( \partial^2\bar h^{KM} )\right]/M^{(n-2)}.
\eea

The contribution from the one loop diagram in Figure \ref{fig:fig3} is deduced by inserting in the momentum space vertex associated with the the action in equation (\ref{div}) an additional factor of $q^{2(n/2-2)}=1-\epsilon \ln(q^2)/2+...$. Without this factor, the contribution to the long range force between strings would vanish. It is the finite nonanalytic part of the effective action, not the divergent part, that is actually responsible for the long range force. The momentum space integral that must be done is then
\be
-{1 \over n-4}\int {d^2 q \over (2 \pi)^2} \exp(i\vec q \cdot \vec a)({\vec q}^2)^{(n/2-2)}= {1 \over 2 \pi a^2}.
\ee
Putting these results together, the contribution from the diagrams in Figure \ref{fig:fig3} to the long range potential energy per unit length between strings is
\bea
\Delta U/L&=&{\tau_ 1\tau_2 \over 16 \pi^3 a^2M^4}\left({1 \over 120}\left[-2+2-{1 \over 2}\right]+{7 \over 20} \left[-{1 \over 2}+1-{1 \over 2}-{1 \over 4}-{1 \over 4}+{1 \over 2}-{1 \over 4}\right]\right) \nonumber  \\
&=&{\tau_ 1\tau_2 \over 16 \pi^3 a^2M^4}\left(-{11 \over 120}\right).
\eea
The successive terms in the square brackets are the contributions of the corresponding terms in the square brackets of equations (\ref{r2}) and (\ref{ri2}).

\begin{figure}
\PSbox{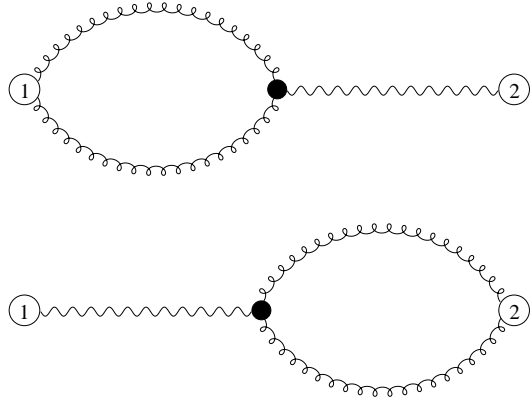 hoffset=120 voffset=10 hscale=35
vscale=35}{6.8in}{2.5in}
\caption{Contribution to the one loop potential that arises from the three-graviton vertex from the Einstein-Hilbert action.}
\label{fig:fig4}
\end{figure}

Next we consider the contribution to the long range force from the Feynman diagrams in Figure \ref{fig:fig4}. For this we need the bulk three-graviton vertex from the Einstein-Hilbert term. It comes from expanding the action in equation (\ref{bulk}) to cubic order in $h$, yielding
\be
S_{3h}=-{2 \over M^{(n/2-1)}}\int d^n x~ {\cal L}_1 +{ \cal L}_2 + {\cal L}_3,
\ee
where ${\cal L}_i$ is the part that comes from expanding the curvature tensor to order $i$ in $h$. Explicitly,
\be
{\cal L}_1={1 \over 8} h^A_A h^B_B \partial^2 h^L_L -{1 \over 8} h^A_A h^B_B \partial_K \partial_N h^{KN} -{1\over 4} h^B_A h^A_B \partial ^2 h^L_L +{1\over 4} h^B_A h^A_B \partial _K \partial_N h^{KN},
\ee
\bea
&&{\cal L}_2 ={1 \over 2}h^A_A (\partial_L h^{EL})(\partial_M h^M_E)-{1 \over 2} h^A_A (\partial_L h^{EL}) (\partial_E h^M_M)  +{1 \over 8} h^A_A (\partial^E h^L_L) (\partial_E h^M_M) \nonumber \\
&&-{3 \over 8} h^A_A (\partial_L h^{EM})(\partial^L h_{EM}) +{1 \over 4} h^A_A (\partial_M h^E_L)(\partial^L h^M_E) -{1 \over 2}h^{NL} h^A_A \partial^2 h_{NL}+h^{NL}h^A_A\partial_M \partial_L h^M_N \nonumber \\
&&-{1 \over 2} h^{NL}h^A_A \partial_N \partial_L h^M_M ,
\eea
and
\bea
&&{\cal L}_3 =-h^G_E(\partial^L h^E_L)(\partial_M h^M_G)+h^G_E (\partial_L h^{LE})(\partial_G h^M_M)-{1 \over 4}h^G_E(\partial^E h^L_L)(\partial_G h^M_M)\nonumber \\
&&-{1 \over 2}h^G_E(\partial_L h^{EM})(\partial_M h^L_G)+{3 \over 2}h^G_E(\partial_L h^{EM})(\partial^L h_{MG})-h^G_E(\partial_L h^{EM})(\partial_G h^L_M)\nonumber \\
&&+{3 \over 4}h^G_E(\partial^E h_{ML})(\partial_G h^{ML})-2h^G_E(\partial_G h^{LE})(\partial_M h^M_L) 
+h^G_E(\partial_G h^{LE})(\partial_L h^M_M)\nonumber \\
&&+h^G_E(\partial^L h^E_G)(\partial_M h^M_L)-{1 \over 2}h^G_E(\partial^L h^E_G)(\partial_L h^M_M) +h^N_Bh^{BL} \partial^2 h_{LN} -2 h^N_B h^{BL}\partial_M \partial_L h^M_N \nonumber \\
&&+h^N_B h^{BL}\partial_N \partial_L h^M_M +h^{NL}h^{MK}\partial_K \partial_M h_{NL}-h^{NL}h^{MK}\partial_K \partial_L h_{NM}.
\eea 
The integral needed to compute Figure \ref{fig:fig4} is
\be 
\int d^2 x_1 d^2 x_2 d^4 x (\partial_x^2 D(x_1-x))D(x-x_2)^2={LT \over 16 \pi^3 a^2},
\ee
and we find that it gives the following contribution to the gravitational potential per unit length:
\bea
\label{r3} 
&&\Delta U/L={\tau_ 1\tau_2 \over 16 \pi^3 a^2M^4}\left( \left[0+0+1-{1 \over 2}\right]+\left[-{1 \over 6}+0+0+{3 \over 4}-{1 \over 12}+0+{1 \over 3}+0\right] \right.\nonumber \\
&&+\left. \left[0+0+0+0+0+0-{1 \over 2}+0+0+1-1+0+0+0-{1 \over 3}+0\right ]   \right),
\eea
In equation (\ref{r3}) the three square brackets contain the contributions from the three Lagrange densities ${\cal L}_1$, ${\cal L}_2$, and ${\cal L}_3$ respectively, and each successive term in these square brackets represents the contribution of the corresponding term in the Lagrange density. Summing them up, we get
\be
\Delta U/L={\tau_ 1\tau_2 \over 16 \pi^3 a^2M^4}\left( {1 \over 2} \right),
\ee
for the contribution from diagrams which contain a three graviton vertex from the Einstein Hilbert action. 

\begin{figure}
\PSbox{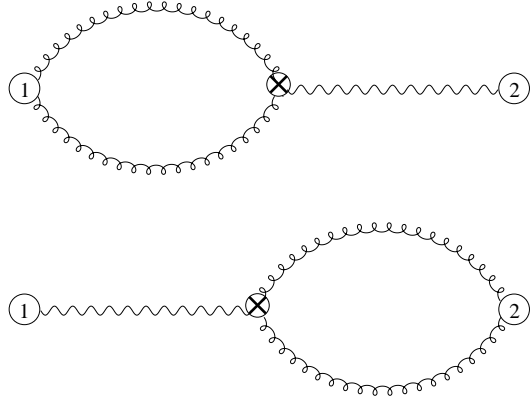 hoffset=120 voffset=10 hscale=35
vscale=35}{6.8in}{2.5in}
\caption{One loop contribution to the potential that arises from the three-graviton vertex from gauge fixing.}
\label{fig:fig5}
\end{figure}

In the background field gauge, the gauge fixing term also contributes to the $\bar h \tilde h \tilde h$ vertex. Using the definition $\bar g_{AB}= \eta_{AB} +\bar h_{AB}/M^{(n/2-1)}$, the gauge fixing term is \cite{hv}
\be
\label{gf}
S_{gf} =- \int d^n x \sqrt{ \bar g}\left(D_N \tilde h^N_M -{1 \over 2} D_M \tilde h^L_L\right)\left(D_S \tilde h^{MS} -{1 \over 2} D^M \tilde h^S_S \right),
\ee 
where in equation (\ref{gf}) indices are raised and lowered with the classical metric $\bar g$ and the covariant derivative is with respect to this metric. Expanding to linear order in $\bar h$, the above becomes
\be
\label{gf1}
S_{gf} =-\int d^n x {1 \over 2M^{(n/2-1)}} \left(\bar h^A_A (\partial_N \tilde h_M^N)(\partial_S \tilde h^{MS})+...\right)+...,
\ee 
where the ellipses in the brackets denote other terms linear in $\bar h$, and the ellipses outside of the brackets denote terms higher order in $\bar h$. Only the term explicitly displayed in equation (\ref{gf1}) contributes at one loop; the other terms linear in $\bar h$ (represented by the ellipses inside the brackets) each give zero. We find that the contribution of Figure \ref{fig:fig5} to the long range potential is

\be
\Delta U/L={\tau_ 1\tau_2 \over 16 \pi^3 a^2M^4}\left( -{1 \over 12} \right).
\ee

All other possible one loop contributions vanish. For example, there is a cubic coupling of $h_{\mu \nu}$ on the brane from expanding the induced metric to that order. The one loop graph formed from this coupling  vanishes since $P^{\alpha}_{\alpha},^{\beta}_{\lambda}=0$. 


So far we have not included the degrees of freedom that correspond to transverse fluctuations of the strings. However, they must exist, by reparametrization invariance and general covariance. These fluctuations are characterized by scalar fields $\phi^a_{(i)}$ which are localized on the world-sheet of string $i$. The terms in the string actions (\ref{brane}) involving the fields $\phi^a_{(i)}$ are deduced from the dependence of the induced metric\footnote{See, for example, \cite{s1}.} on them, 
\be
g^{(i)}_{\mu \nu}= g_{\mu \nu} + g_{a b}(\partial_{\mu} \phi_{(i)}^a)(\partial_{\nu} \phi_{(i)}^b).
\ee
Expanding the square root of the determinant of the above induced metric yields a coupling of $h_{ab}$ to the scalar fields. However, the graph with a $\phi$ loop vanishes in dimensional regularization since it is proportional to $\int d^{(n-2)}k=0$.


Summing the various one loop contributions to gravitational potential energy between strings gives
\be
U/L={\tau_ 1\tau_2 \over 16 \pi^3 a^2M^4}\left( {3 \over 40} \right)={24G_N^2\tau_1\tau_2 \over5 \pi a^2}.
\ee
The above equation is the main result of this paper. It gives a repulsive gravitational force between the strings at large distances. 

From the effective field theory point of view, it is possible that the tree level effects of higher dimension operators are of the same size as the one loop pieces we have calculated, but this turns out not to be the case.  Nontrivial operators localized on the string world sheet with fewer than two derivatives are forbidden by general covariance. Furthermore, we know that many operators do not contribute to the long range force. Consider, for example, adding to the string world-sheet actions the following two-derivative term:
\be
\delta S_i = \lambda_i \int d^4 x \sqrt{ g^{(i)}}~ R~ \delta^{(2)} (\vec x-\vec x_i),
\ee
where the $\lambda_i$ are dimensionless couplings. Classically, there is no  contribution to the long range force between the branes linear in these couplings. At this order, it gives only local effects proportional to $\delta^{(2)}(\vec a)$ or derivatives of this delta function. Similar remarks hold for operators in the bulk that are quadratic in the curvature tensor. We will not attempt a complete analysis of the tree level effects from higher dimension operators; however, there is no tree level contribution to the potential that is as important at large $a$ as the one loop piece we have calculated.

Effects similar to what we have computed occur for $p$-branes in a space-times of dimension $p+3$. Assuming that gravity is the only massless degree of freedom in the bulk, there will be a long range contribution to the potential per unit $p$-brane volume proportional to $G_N^2 \tau_1 \tau_2/a^{p+1}$ from one loop quantum effects. If there are other massless degrees of freedom in the bulk, these will also contribute to the long range force. Consider, for example, a scalar field theory with space-time dimension $p+3$ and two parallel $p$-branes. Neglecting gravity, the action for this system is taken to be
\be
\label{act} 
S=S_b+S_1+S_2,
\ee
where the bulk action, $S_b$, comes from the massless Klein Gordon theory:
\be
S_b=-{1 \over 2}\int d^{p+3}x~\partial_M \chi \partial^M \chi,
\ee
and the brane actions are
\be
S_i=-{\lambda_i \over 2}\int d^{p+3}x~ \chi^2 \delta^{(2)} (\vec x-\vec x_i).
\ee

\begin{figure} 
\PSbox{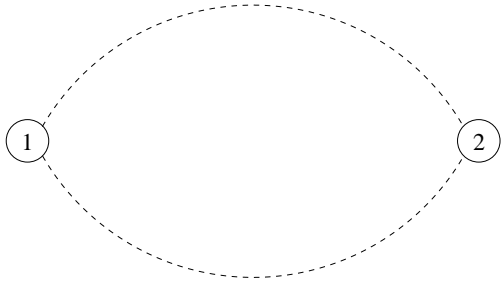 hoffset=100 voffset=10 hscale=40
vscale=40}{6.8in}{2.0in}
\caption{One loop contribution to the potential from massless bulk scalar with brane mass terms.}
\label{fig:fig8}
\end{figure}

Because of the $\chi \rightarrow - \chi$ symmetry there is no tree level force between the branes from $\chi$ exchange. Assuming that the couplings $\lambda_i$ are small and neglecting effects higher order in these coupling constants, the one loop diagram in Figure \ref{fig:fig8} gives the long range potential\footnote{For work in string theory 
on the force between branes see \cite{bs}.}
\be
\label{spot}
U/V=-{\lambda_1 \lambda_2  \Gamma({p \over 2}+{1 \over 2})^2 \over a^{p+1}p 2^{p+4} \pi^{({p \over 2}+2)}\Gamma({p\over 2})}.
\ee
If the couplings $\lambda_1$ and $\lambda_2$ have opposite signs, this potential is repulsive.\footnote{There may be an instability of the background $\chi=0$ configuration in this case.  Since this is only a ``toy model,'' we have not explored this issue further.}  It can be natural for the scalar to have brane mass terms but no bulk mass. For example, $\chi$ could be the Goldstone boson associated with a global symmetry that is spontaneously broken in the bulk but explicitly broken on the branes. At higher order the couplings $\lambda_i$ become subtraction point dependent \cite{gw}.

Let's focus on the case of three-branes in six dimensions. If the two dimensions perpendicular to the branes are compact but large extra dimensions of the type that has been suggested to be related to the hierarchy puzzle \cite{add}, then the potential in equation (\ref{spot}) has the right form to be suitable for cosmological quintessence\footnote{ This is similar to the proposal in \cite{abrs}.}. The scalar field has mass dimension two, so the parameters $\lambda_i$ are dimensionless. The separation between the branes is related to the scalar fields that characterize the brane fluctuations. Assuming the two compact extra dimensions are flat,\footnote{The physics that determines the size of the compact two extra dimensions is assumed to be unrelated to the potential generated by $\chi$ loops.} the action for the scalar fields that characterize the fluctuations of the 3-brane world-sheets is
\be
S_{fluct}=-\tau_1 \int d^4 x_1 {1 \over 2}\partial_{\mu}\phi^a_{(1)} \partial^{\mu} \phi^a_{(1)}-\tau_2 \int d^4 x_2 {1 \over 2}\partial_{\mu}\phi^a_{(2)} \partial^{\mu} \phi^a_{(2)} +....~.
\ee
The repeated index $a$, which takes on the values $1,2$, is summed over. From the four dimensional effective field theory point of view, this action becomes
\be
S^{eff}_{4dim}=-\int d^4x~ { \tau_1+\tau_2 \over 2}\partial_{\mu} \phi^a_{cm} \partial^{\mu} \phi^a_{cm}+{\tau_r \over 2} \partial_{\mu} \phi^a_{rel}\partial^{\mu} \phi^a_{rel}+...,
\ee
where $\tau_r=\tau_1 \tau_2/(\tau_1+\tau_2)$ is the reduced tension, $\phi^a_{rel}= \phi^a_{(1)}- \phi^a_{(2)}$ and $\phi^a_{cm}=(\tau_1 \phi^a_{(1)}+\tau_2 \phi^a_{(2)})/(\tau_1+\tau_2)$. The separation between the branes $\vec a $ is the vacuum expectation value of $\vec \phi_{rel}$, and the canonically normalized four dimensional field associated with the separation between the branes is $\vec \phi=\sqrt \tau_r \vec \phi_{rel}$. The potential for this scalar field is of the form $ U/V \sim -\lambda_1 \lambda_2 \tau_r^2/  (\vec \phi ^2)^2 \sim M_W^8/(\vec \phi^2)^2$, when the brane tensions are of order the weak scale\footnote{With tensions this large it is no longer a good approximation to neglect the deficit angles associated with these three-branes.}. In cosmological quintessence the scalar field today is of order the Planck mass; this corresponds to a separation between branes of order the size of the compact space ({\it i.e.} of order a millimeter). Clearly, the impact of the physics that stabilizes the compact dimensions  \cite{stab} has to be taken into account before the true physical significance of this potential can be ascertained.

We thank R. Sundrum and A. Lewandowski for some useful discussions. This work was supported in part by the department of energy under grant DE-FG03-92-ER-40701.

\end{document}